\documentclass[11pt]{article}
\usepackage{cospar}
\usepackage{url}

\usepackage{graphicx}

\newcommand\go{\mathrel{\raise.3ex\hbox{$>$}\mkern-14mu
             \lower0.6ex\hbox{$\sim$}}}
\newcommand\lo{\mathrel{\raise.3ex\hbox{$<$}\mkern-14mu
             \lower0.6ex\hbox{$\sim$}}}
\newcommand{\Teff}{T_{\rm eff}}
\newcommand{\veck}{\hat {\bf k}}
\newcommand{\vecB}{\hat {\bf B}}
\newcommand{\vecE}{{\bf E}}

\title{ATMOSPHERES OF MAGNETIZED NEUTRON STARS: VACUUM POLARIZATION
AND PARTIALLY IONIZED MODELS}

\author{Wynn C.G. Ho\address{Center for Radiophysics and Space Research,
    Department of Astronomy, Cornell University, Ithaca, NY 14853, USA},
        Dong Lai$^{1}$,
        Alexander Y. Potekhin\address{Ioffe Physico-Technical Institute, 
        Politekhnicheskaya 26, 194021, St. Petersburg, Russia},
        and
        Gilles Chabrier\address{CRAL (UMR CNRS No. 5574),
        Ecole Normal Sup\'{e}rieure de Lyon, 69364 Lyon Cedex 07, France}}

\begin{document}

\maketitle

\begin{abstract}
We construct hydrogen atmosphere models for magnetized neutron stars in
radiative equilibrium with surface fields $B=10^{12}-5\times 10^{14}$~G
and effective temperatures $\Teff\sim\mbox{a few}\times 10^5-10^6$~K
by solving the full radiative transfer equations for both polarization
modes in the magnetized hydrogen plasma.
The atmospheres directly determine the characteristics of thermal
emission from isolated neutron stars.
We study the effects of vacuum polarization and bound atoms on the
atmosphere structure and spectra.
For the lower magnetic field models ($B\sim 10^{12}$~G), the spectral
features due to neutral atoms lie at extreme UV and very soft X-ray
energies and
therefore are not likely to be observed.  However, the continuum
flux is also different from the fully ionized case, especially at
lower energies.
For the higher magnetic field models, we find that vacuum polarization
softens the high energy tail of the thermal spectrum.  We show
that this depression of continuum flux strongly suppresses not only
the proton cyclotron line but also spectral features due to bound species;
therefore spectral lines or features in thermal radiation are more
difficult to observe when the neutron star magnetic field is $\go 10^{14}$~G.
\end{abstract}

\section*{INTRODUCTION}

Thermal radiation from the surface of isolated neutron stars (NSs)
can provide invaluable information on the physical properties and
evolution of NSs.  Such radiation has been detected in various
types of isolated NSs: from radio pulsars (see Becker~2000;
Pavlov et al.~2002) to old and young radio-quiet NSs
(see Treves et al.~2000; Pavlov et al.~2002)
to soft gamma-ray repeaters (SGRs) and anomalous X-ray pulsars (AXPs)
(see Hurley~2000; Israel, Mereghetti, \& Stella~2001; Mereghetti et al.~2002),
which form a potentially new class of NSs
(``magnetars'') endowed with superstrong ($B\go 10^{14}$~G) magnetic
fields (see Thompson \& Duncan~1996; Thompson~2001).
The NS surface emission is
mediated by the thin atmospheric layer (with scale height
$\sim 0.1-10$~cm and density $\sim 0.1-100$~g/cm$^3$) that
covers the stellar surface.  Therefore, to properly interpret
the observations of NS surface emission and to provide
accurate constraints on the physical properties of NSs, it is
important to understand in detail the radiative properties
of NS atmospheres in the presence of strong magnetic fields.

Steady progress has been made over the years in modeling NS atmospheres
(see Pavlov et al.~1995; Ho \& Lai~2001, 2003; Zavlin \& Pavlov~2002
for more detailed references on observations and on previous works
of NS atmosphere modeling).
These atmosphere models have played a valuable role in assessing
the observed spectra of radio pulsars and radio-quiet NSs.
We have studied the H and He atmospheres of NSs with magnetic fields
$B\sim 10^{12}-10^{15}$~G and effective temperatures
$T_{\rm eff}\sim 10^5-10^7$~K (Ho \& Lai~2001, 2003;
Ho et al.~2002).
Here we briefly discuss our atmosphere model and the effects of
vacuum polarization and report on recent work in incorporating hydrogen
bound species in the models to produce partially ionized atmospheres.

\section*{ATMOSPHERE MODEL}

We consider an isolated NS with a plane-parallel
atmosphere.  This is justified since the atmospheric scale
height $H \lo 10$~cm is much less than the NS radius
$R\approx 10$~km.  The atmosphere is composed of pure hydrogen.
A uniform magnetic field ${\bf B}$, aligned perpendicular to
the surface, permeates the atmosphere, and the spectra presented
indicate emission from a local patch of the NS surface with the
indicated magnetic field strength and effective temperature.

In the highly magnetized plasma that characterizes NS atmospheres,
there are two photon polarization modes, and they have very different
radiative opacities.
The two photon modes are the
extraordinary mode (X-mode), which is mostly polarized perpendicular
to the $\veck-\vecB$ plane, and the ordinary mode (O-mode), which is
mostly polarized parallel to the $\veck-\vecB$ plane,
where $\veck$ specifies the direction of photon propagation
and $\vecB$ is the direction of the external magnetic field.
The O-mode has a significant component of its electric field $\vecE$
along $\vecB$ for most directions of propagation, except when $\veck$ is 
nearly parallel to $\vecB$, and therefore the O-mode opacity is close
to the $B=0$ value, while the X-mode opacity is much smaller.

To construct self-consistent atmosphere models requires successive
iterations, where the temperature profile $T(\tau)$ is adjusted from
the previous iteration in order to satisfy radiative equilibrium
($\tau$ is the Thomson depth within the atmosphere).
We solve the radiative transfer equations (RTEs) for the two coupled photon
polarization modes, together with the boundary conditions,
by the finite difference scheme described in Mihalas~(1978).
We use a variation of the Uns\"{o}ld-Lucy
temperature correction method as described in Mihalas~(1978)
but modified to account for full radiative transfer by two propagation
modes in a magnetic medium.
The process of determining the radiation intensity from
the RTE for a given temperature profile, estimating and
applying the temperature correction, and then recalculating
the radiation intensity is repeated until convergence
of the solution is achieved (see Ho \& Lai~2001 for details
of our numerical method).

\section*{VACUUM POLARIZATION}

It is well-known that polarization of the vacuum due to virtual
$e^+e^-$ pairs becomes significant when $B\go B_{\rm Q}$, where
$B_{\rm Q}=m_{\rm e}^2c^3/e\hbar=4.414\times 10^{13}$~G.  Vacuum
polarization modifies the dielectric property of the medium
and the polarization of photon modes, thereby altering the
radiative scattering and absorption opacities 
(see Ho \& Lai~2003 and references therein).
Of particular interest is the ``vacuum resonance'' phenomenon, which
occurs when the effects of the vacuum and plasma on the linear
polarization of the modes cancel each other, giving rise to
a resonance feature in the opacity at  photon energy
$E_{\rm V} \approx 1.02\,(Z/A)(\rho/\mbox{1 g cm$^{-3}$})^{1/2}
(B/\mbox{10$^{14}$ G})^{-1}f(B)$~keV,
where $Z$ and $A$ are the atomic charge and mass of the ion, respectively,
$\rho$ is the density,
and $f(B)$ is a slowly-varying function of $B$ of order unity.

It was shown in Lai \& Ho (2002, 2003) (see also
Gnedin et al.~1978; Pavlov \& Gnedin~1984) that high energy photons
propagating in the atmospheric plasma
can adiabatically convert from one polarization mode into another
at the vacuum resonance density
$\rho_{\rm V} \approx 0.96\,(A/Z)(E/\mbox{1 keV})^2
 (B/\mbox{10$^{14}$ G})^2f(B)^{-2}\mbox{ g cm$^{-3}$}$.
Across the resonance, the orientation of the polarization ellipse
rotates by $90^\circ$, although the helicity does not change
(see Figure~1).
This resonant mode conversion is analogous to the
Mikheyev-Smirnov-Wolfenstein (MSW) effect for neutrino oscillations.
Because the two photon modes have vastly different opacities,
this vacuum-induced mode conversion can significantly affect
radiative transfer in strongly magnetized atmospheres.
Figure~2 illustrates the main effect that vacuum
polarization has on the atmosphere spectrum.
The left side of Figure~2 shows the decoupling densities of the
O-mode and X-mode photons (i.e., the densities of their respective
photospheres) when the vacuum polarization effect is neglected.
The right side of Figure~2 shows that the effective decoupling depths
of the photons are changed when the vacuum resonance lies between
these two photospheres.  We see from Figure~2 that mode conversion
makes the effective decoupling density of X-mode photons (which carry the
bulk of the thermal energy) smaller, thereby depleting the high energy
tail of the spectrum and making the spectrum closer to blackbody (although
the spectrum is still harder than blackbody because of non-grey opacities).
This expectation is borne out in self-consistent atmosphere modeling
presented in Ho \& Lai~2003 and illustrated in Figure~3\footnote{
All the models presented here use the corrected free-free absorption,
which properly accounts for electron-ion collisions
(Potekhin \& Chabrier~2003; see also Ho et al.~2002).
}
for the case of $B=5\times 10^{14}$~G and $\Teff=5\times 10^6$~K.

Another important effect of vacuum polarization on the spectrum, first
noted in Ho \& Lai~2003 and illustrated in Figure~3, is the suppression
of proton cyclotron lines
(and other spectral lines; see below).  The physical origin for such line
suppression is related to the depletion of continuum flux; this depletion
makes the decoupling depths inside and outside the line similar.

\begin{center}
\begin{table}[t]
\begin{minipage}{75mm}
\includegraphics[width=75mm]{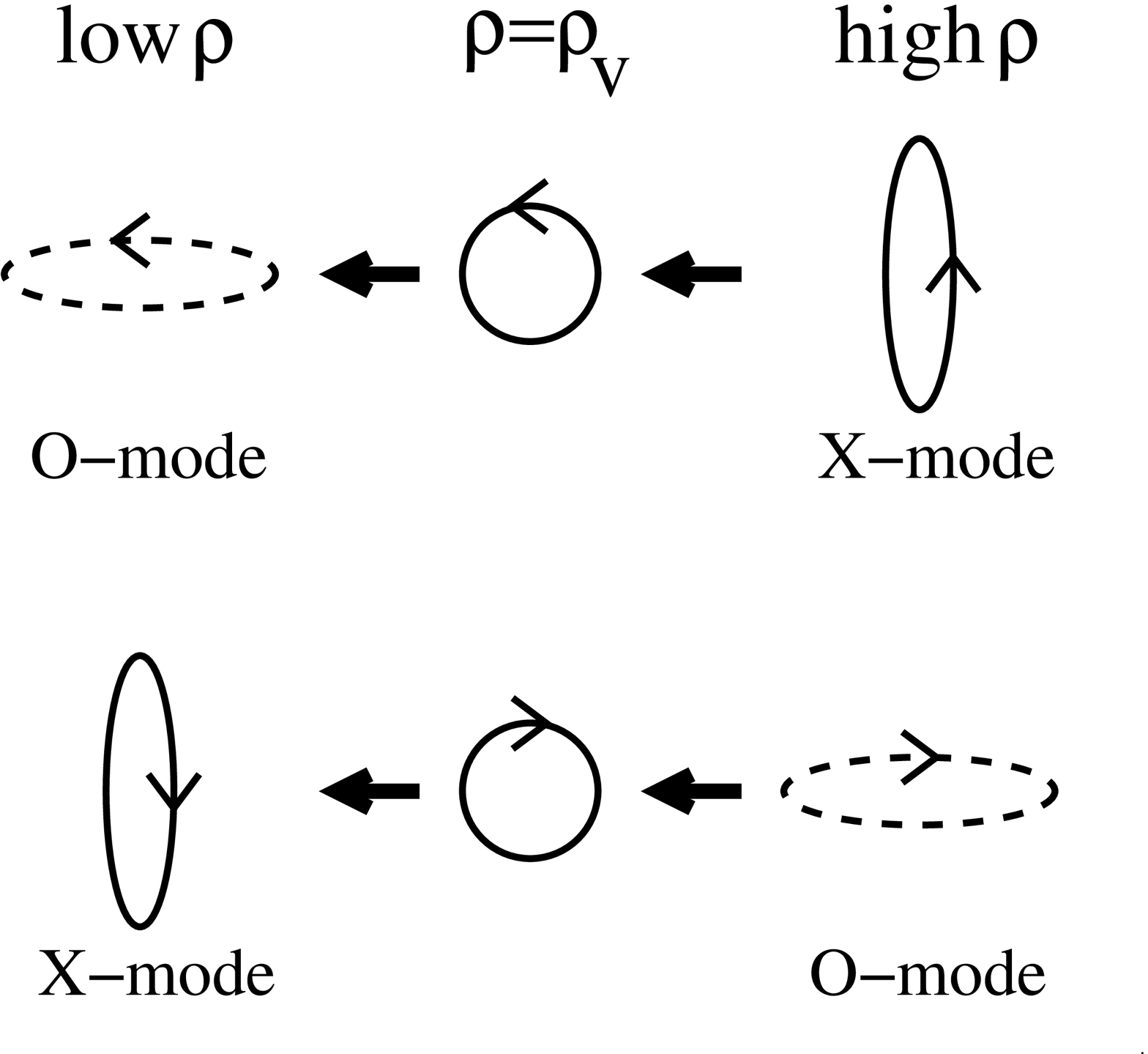}

{\sf Fig. 1. A schematic diagram illustrating mode conversion due to
vacuum polarization: As an X-mode (O-mode) photon at high density
traverses the vacuum resonance density $\rho_V$, it becomes an O-mode
(X-mode) photon at low density if the adiabatic condition is
satisfied.  In this evolution, the polarization ellipse rotates $90^\circ$,
and the photon opacity changes significantly.}
\end{minipage}
\hfil\hspace{\fill}
\begin{minipage}{75mm}
\includegraphics[width=75mm]{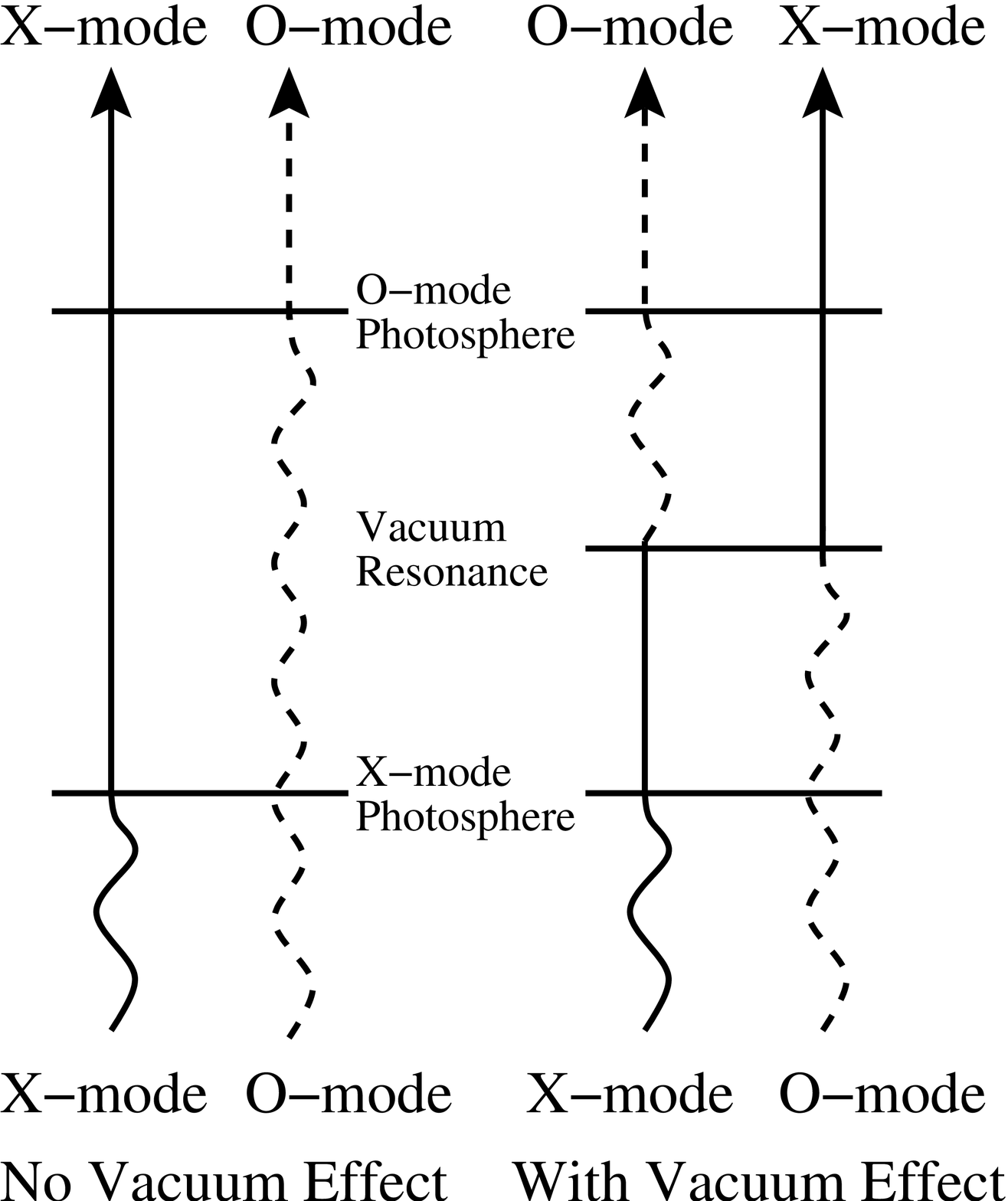}
\\

{\sf Fig. 2. A schematic diagram illustrating how vacuum polarization-induced 
mode conversion affects radiative transfer in a magnetar atmosphere.
When the vacuum polarization effect is turned off, the X-mode 
photosphere (where optical depth $\sim 1$) lies deeper than the O-mode. 
With the vacuum polarization effect included, the X-mode effectively
decouples (emerges) from the atmosphere at the vacuum resonance,
which lies at a lower density than the (original) X-mode photosphere.}
\end{minipage}
\end{table}
\end{center}

\section*{PARTIALLY IONIZED ATMOSPHERES}

A major complication when incorporating bound species in the atmosphere
models arises from the strong coupling between the center-of-mass motion
of the atom and the internal atomic structure.
Potekhin \& Chabrier~(2003) have constructed the first thermodynamically
consistent equation of state and opacity models of partially ionized
hydrogen plasmas in strong magnetic fields.
From the work of Potekhin \& Chabrier~(2003), we obtain tables for
the equation of state and absorption opacities of hydrogen, including
the bound states.
We interpolate between the table values to obtain densities
and absorption opacities corresponding to the numerical grid of
our models.\footnote{
Note that we calculate the polarization vectors of the medium
assuming the medium is fully ionized since the abundance of bound
species is very low for the cases considered; this is not strictly
correct (see Bulik \& Pavlov~1996 for the case where the atmosphere
is completely neutral).}
Thus we are able to obtain partially ionized hydrogen atmosphere models
(Ho et al.~2002).

Figure~4 shows the spectra of hydrogen atmospheres with $B=10^{12}$~G
and $\Teff=5\times 10^5$~K, along with the non-magnetic hydrogen
atmosphere model at the same $\Teff$ and the blackbody spectrum
with $T=5\times 10^5$~K.  The proton cyclotron line at 6.3~eV
is clear in both the fully ionized and partially ionized spectra.\footnote{
The width of the proton cyclotron line in all the partially ionized
models is due to the finite energy grid of the models and not an
indication of the true line width; the true width is narrower.}
It is clear from Figure~4
that spectral features due to bound species at this field strength
lie within the extreme UV
($E\go 50$~eV) to very soft X-ray ($E\lo 0.2$~keV) regime and
therefore are likely to be unobservable due to interstellar absorption.
However, the effect of bound species on the temperature profile
of the atmosphere and the continuum flux is significant.  In
particular, the optical flux is higher for the partially ionized
atmospheres compared to the fully ionized atmospheres for the
same $\Teff$.  On the other hand, the
partially and fully ionized models both yield very similar (neglecting
the spectral features) X-ray flux for a given effective temperature,
with the partially ionized model fluxes slightly lower than those
from the fully ionized models; thus fitting only the high energy flux
with fully ionized models would yield fairly accurate NS temperatures.

Figure~5 compares the partially and fully ionized models
(neglecting vacuum polarization) of hydrogen atmospheres with
$B=5\times 10^{14}$~G and $\Teff=5\times 10^6$~K.
Neglecting vacuum polarization allows us to distinguish
features due to neutral atoms in the atmosphere;
broad absorption features due to bound-free and bound-bound
transitions at $E\sim 0.76$~keV and 4~keV, respectively, are apparent
(the latter is from blending of two bound-bound transitions
at 3.4~keV and 6.5~keV).
The continuum flux between the models is not
significantly different, and as noted in Ho \& Lai~(2001, 2003)
and Zane et al.~(2001),
the proton cyclotron line is very broad when vacuum polarization
is neglected.
Figure~6 compares the models
with vacuum polarization (complete mode conversion; see Ho \& Lai~2003).
We see that vacuum polarization softens the high energy tail of the
spectrum in the partially ionized model, like the fully ionized model.
In addition, we see that vacuum polarization also suppresses spectral
features due to the bound species of hydrogen.  The reduced width
of the proton cyclotron line and spectral features associated with
bound transitions makes these features difficult to observe with
current X-ray detectors, and
we suggest that the absence of lines in the observed spectra of several AXPs
(Patel et al.~2001; Juett et al.~2002; Tiengo et al.~2002) may be an
indication of the vacuum polarization effect at work in these systems.

\begin{center}
\begin{table}[t]
\begin{minipage}{75mm}
\includegraphics[width=75mm]{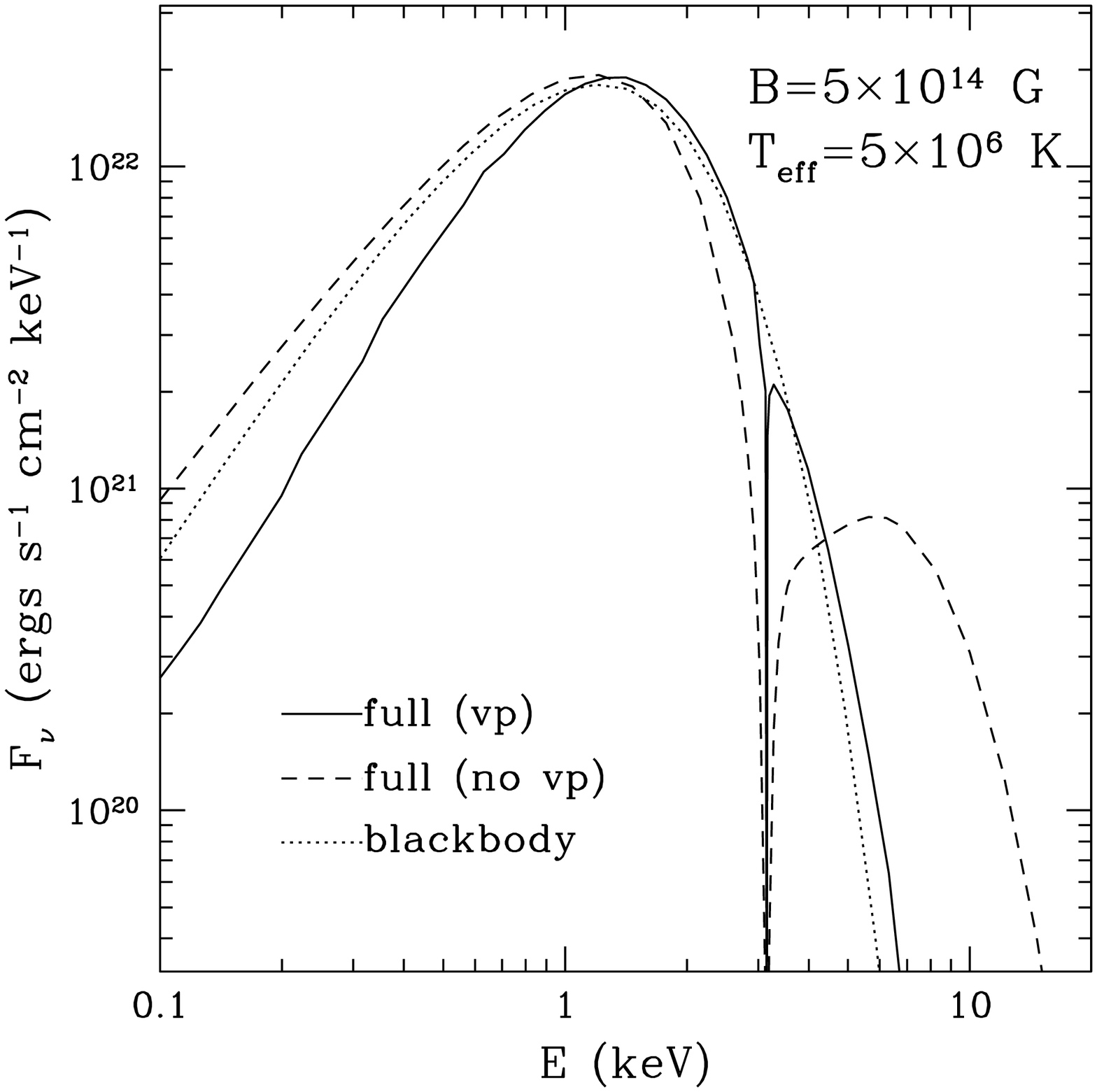}

{\sf Fig. 3. Spectra of fully ionized hydrogen atmospheres with
$B=5\times 10^{14}$~G and $\Teff=5\times 10^6$~K.  The solid line
is for an atmosphere which includes vacuum polarization, the
dashed line is for  an atmosphere which neglects vacuum polarization,
and the dotted line is for a blackbody with $T=5\times 10^6$~K.}
\end{minipage}
\hfil\hspace{\fill}
\begin{minipage}{75mm}
\includegraphics[width=75mm]{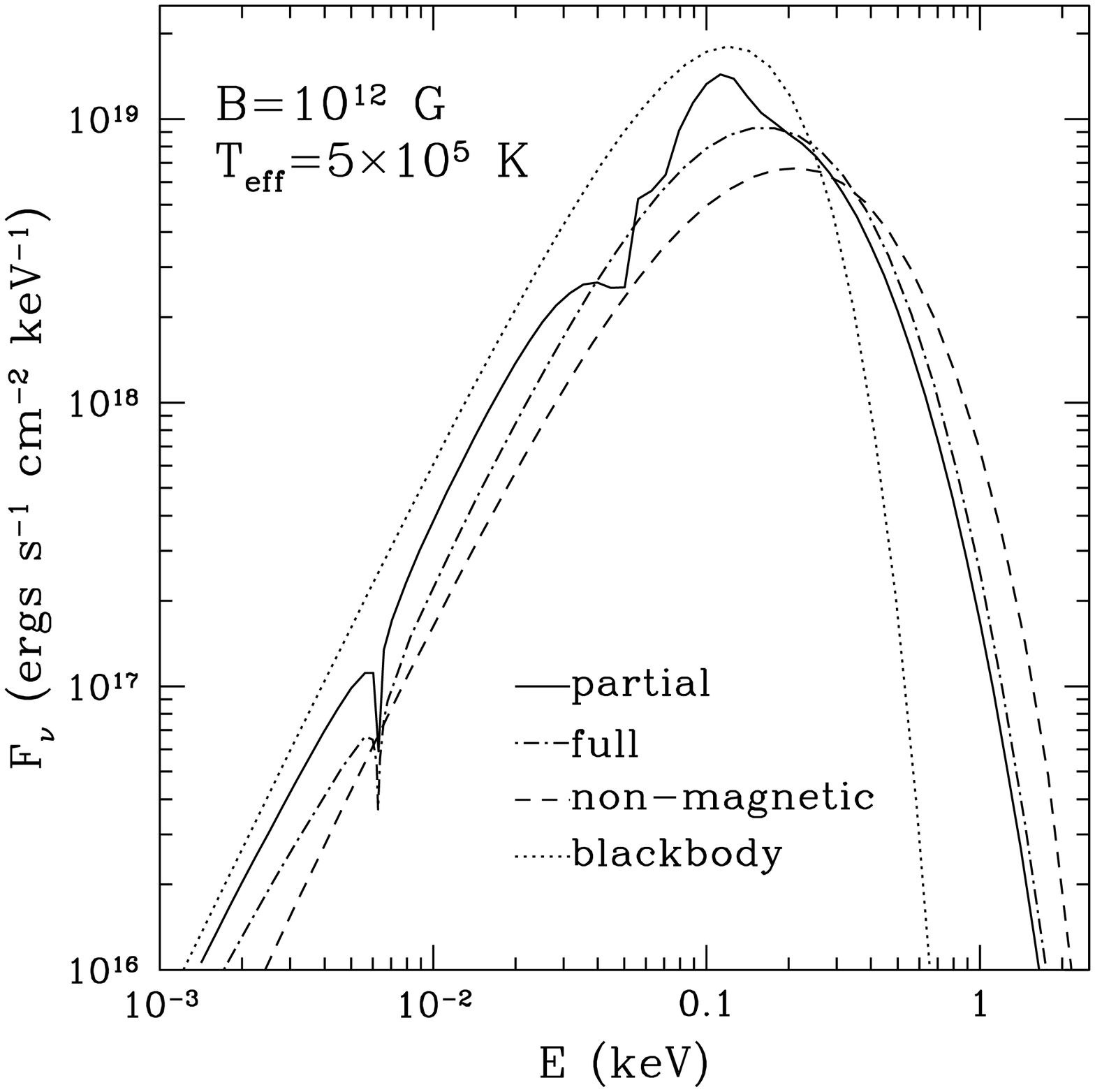}

{\sf Fig. 4. Spectra of hydrogen atmospheres with $B=10^{12}$~G and
$\Teff=5\times 10^5$~K.  The solid line is for a partially ionized
atmosphere, the dot-dashed line is for a fully ionized atmosphere,
the dashed line is for a fully ionized non-magnetic atmosphere,
and the dotted line is for a blackbody with $T=5\times 10^5$~K.}
\end{minipage}
\end{table}
\end{center}

\begin{center}
\begin{table}[t]
\begin{minipage}{75mm}
\includegraphics[width=75mm]{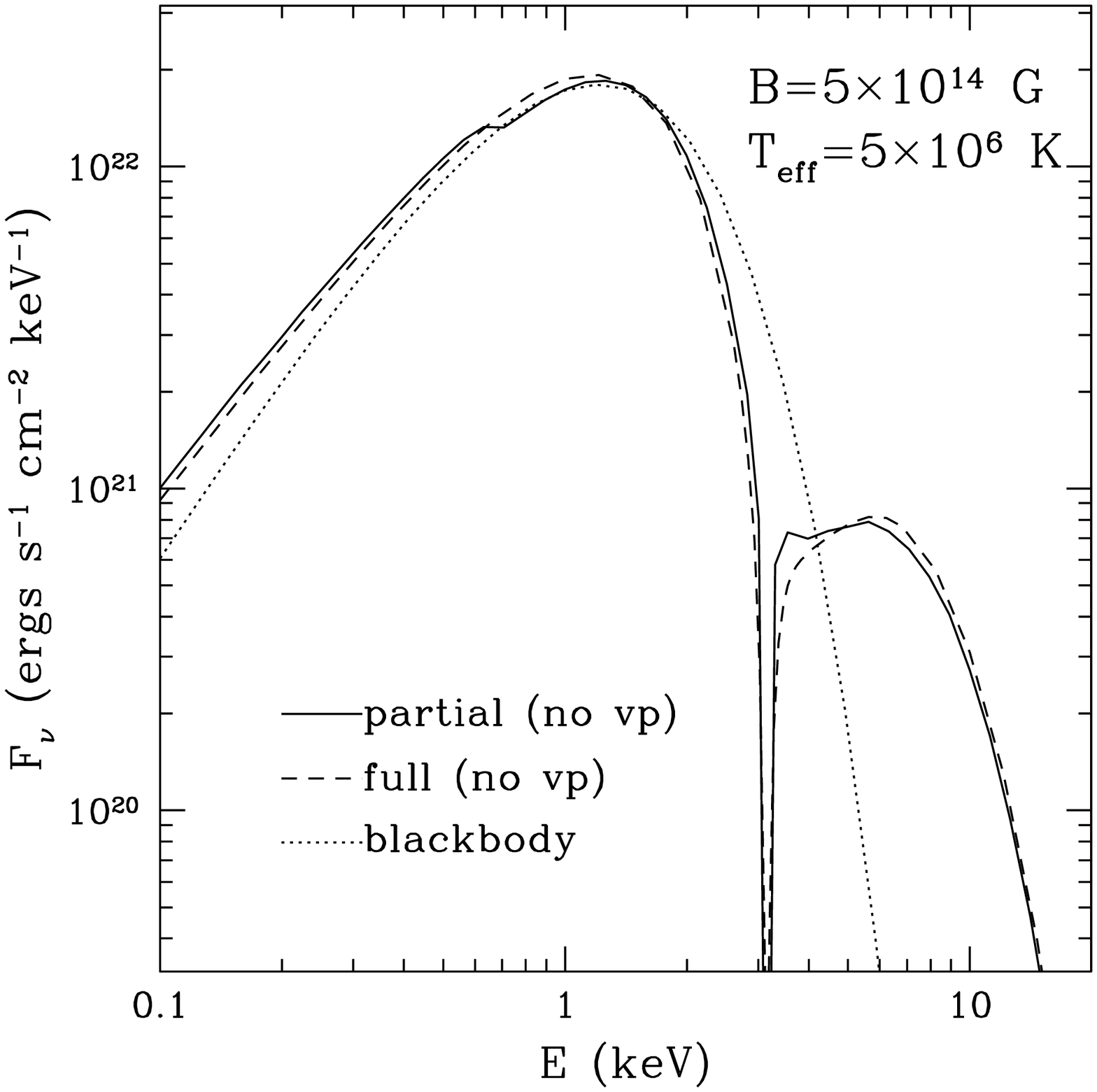}

{\sf Fig. 5. Spectra of hydrogen atmospheres with $B=5\times 10^{14}$~G,
$\Teff=5\times 10^6$~K, and neglects vacuum polarization.
The solid line is for a partially ionized atmosphere,
the dashed line is for a fully ionized atmosphere,
and the dotted line is for a blackbody with $T=5\times 10^6$~K.}
\end{minipage}
\hfil\hspace{\fill}
\begin{minipage}{75mm}
\includegraphics[width=75mm]{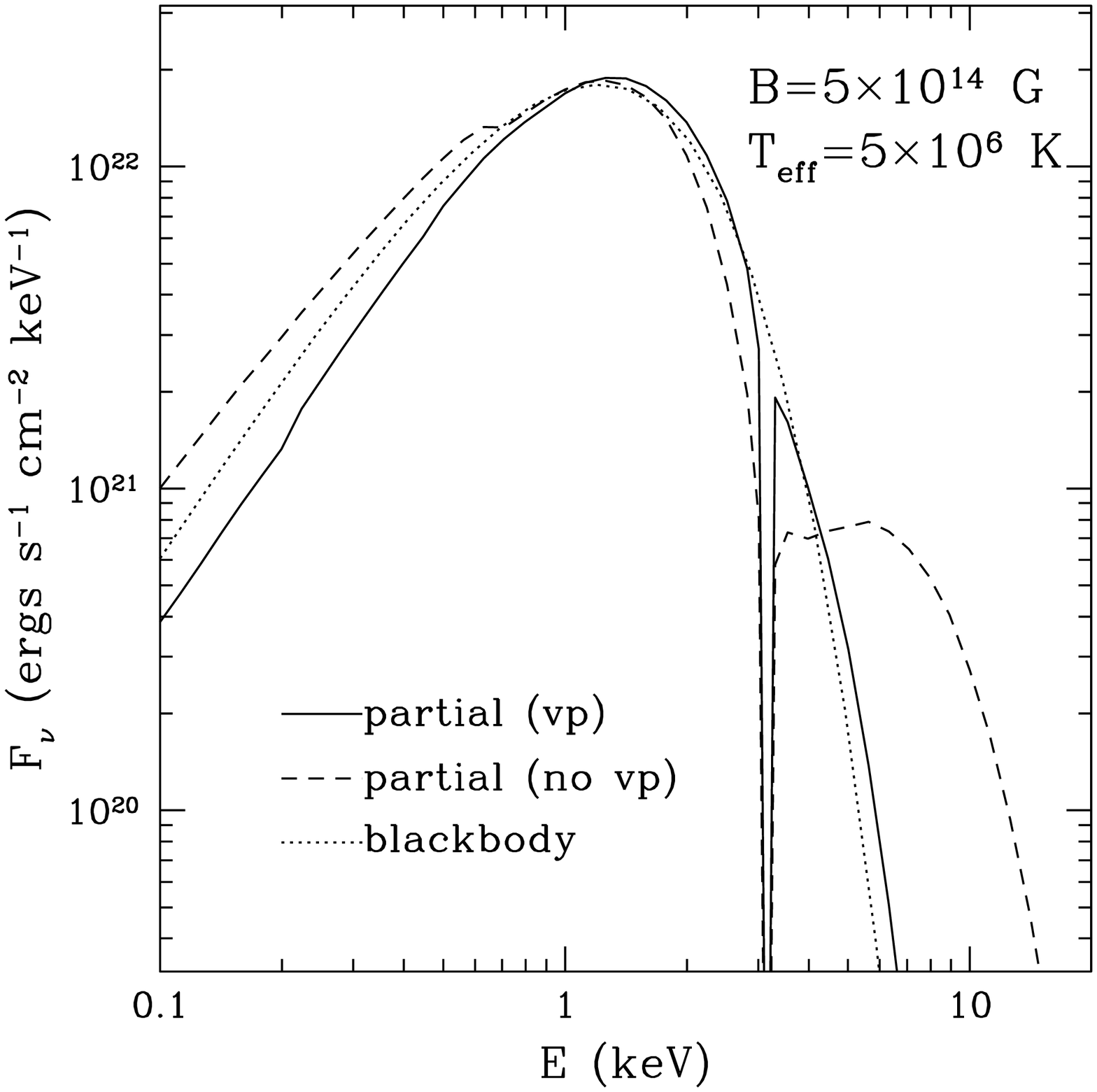}

{\sf Fig. 6. Spectra of partially ionized hydrogen atmospheres with
$B=5\times 10^{14}$~G and $\Teff=5\times 10^6$~K.  The solid line
is for an atmosphere which includes vacuum polarization, the
dashed line is for  an atmosphere which neglects vacuum polarization,
and the dotted line is for a blackbody with $T=5\times 10^6$~K.}
\end{minipage}
\end{table}
\end{center}

\section*{ACKNOWLEDGMENTS}

We are grateful to the Cornell Hewitt Computer Laboratory for the
use of their facilities.
This work is supported in part by NASA grant
NAG~5-12034 and NSF grant AST~9986740.
D.L. is also supported by a fellowship from the A.P. Sloan Foundation.
The work of A.Y.P. is supported by RFBR grant 02-02-17668

\end{document}